\documentclass[times, 10pt,twocolumn]{article}%
\usepackage{latex8}
\usepackage{times}
\usepackage{amsmath}
\usepackage{amsfonts}
\usepackage{amssymb}
\usepackage{graphicx}%
\begin{document}

\title{Lattice Gas Cellular Automata for Computational Fluid Animation}
\author{Gilson A. Giraldi, Adilson V. Xavier, Antonio L. Apolinario Jr, Paulo S. Rodrigues\\National Laboratory of Scientific Computing\\Ave Get\'{u}lio Vargas, 333, 25651-075, Petr\'{o}polis, RJ, Brasil\\\{gilson,adilson,alopes,pssr\}@lncc.br\\}
\maketitle

\begin{abstract}
The past two decades showed a rapid growing of physically-based modeling of
fluids for computer graphics applications. In this area, a common top down
approach is to model the fluid dynamics by Navier-Stokes equations and apply a
numerical techniques such as Finite Differences or Finite Elements for the
simulation. In this paper we focus on fluid modeling through Lattice Gas
Cellular Automata (LGCA) for computer graphics applications. LGCA are discrete
models based on point particles that move on a lattice, according to suitable
and simple rules in order to mimic a fully molecular dynamics. By
Chapman-Enskog expansion, a known multiscale technique in this area, it can be
demonstrated that the Navier-Stokes model can be reproduced by the LGCA
technique. Thus, with LGCA we get a fluid model that does not require solution
of complicated equations. Therefore, we combine the advantage of the low
computational cost of LGCA and its ability to mimic the realistic fluid
dynamics to develop a new animating framework for computer graphics
applications. In this work, we discuss the theoretical elements of our
proposal and show experimental results.

\end{abstract}

\thispagestyle{empty}

\Section{Introduction}

Physically-based techniques for the animation of natural elements like fluids
(gas or liquids), elastic, plastic and melting objects, among others, have
taken the attention of the computer graphics community \cite{1028542}. The
motivation for such interest rely in the potential applications of these
methods and in the complexity and beauty of the natural phenomena that are
involved \cite{311549, desbrun96}. In particular, techniques in the field of
Computational Fluid Dynamics (CFD) have been applied for fluid animation in
applications such as virtual surgery simulators, computer games and visual
effects \cite{SIGGR2004,Nguyen}.

In this paper we focus on physically-based fluid animation for computer
graphics applications (see \cite{SIGGR2004} and references therein).
Basically, the works in this area fall in to two categories: Realistic fluid
and Interactive, or Real-Time, fluid animation. The former is more suitable
for the special effects industry \cite{Nguyen} while the later is appropriate
for interactive applications like computer games and virtual surgery
\cite{Stam2003,1011149}. The work \cite{Foster-Metaxas-1997} is a remarkable
one in this area which includes fluid equations and numerical technique
\cite{Hirsch1988}, shortly Computational Fluid Dynamics (CFD), and scientific
visualization methods \cite{rosemblum94}. The literature of this field reports
gas \cite{Foster-Metaxas-1997,311549} and water simulations \cite{muller03},
interaction between liquids and deformable solids
\cite{Muller2004SolidsFluids}, and others \cite{Thalmann1991, SIGGR2004}.

A majority of fluid animation methods in computer graphics use 2D/3D mesh
based approaches that are mathematically motivated by the Eulerian methods of
Finite Element (FE) and Finite Difference (FD), in conjunction with
Navier-Stokes equations of fluids \cite{Hirsch1988}. These works are based on
a top down viewpoint of the nature: the fluid is considered as a continuous
system subjected to Newton's and conservation Laws as well as state equations
connecting the macroscopic variables of pressure $P,$ density $\rho$ and
temperature $T$.

In this paper, we change the viewpoint to the bottom up model of the Lattice
Gas Cellular Automata (LGCA) \cite{Frisch87}. These are discrete models based
on point particles that move on a lattice, according to suitable and simple
rules in order to mimic a fully molecular dynamics. Particles can only move
along the edges of the lattice and their interactions are based on simple
collision rules. There is an exclusion principle that limits to one the number
of particles that enter a given site (lattice node) in a given direction of
motion. Such framework needs low computational resources for both the memory
allocation and the computation itself. Such models have been applied for
scientific application in two-phase flows description (gas-liquid systems, for
example), numerical simulation of bubble flows \cite{Inamuro2004}, among
others. Besides, Wolfram \cite{Wolfram1996} has studied the computational and
thermodynamics aspects of these models for fluid modeling.

In this paper we focus on fluid modeling through Lattice Gas Cellular Automata
(LGCA) for computer graphics applications. Specifically we take a special
LCGA, introduced by Frisch, Hasslacher and Pomeau, known as FHP model, and
show its capabilities for computer graphics applications. By Chapman-Enskog
expansion, a known multiscale technique in this area, it can be demonstrated
that the Navier-Stokes model can be reproduced by FHP technique. However,
there is no need to solve Partial Differential Equations (PDEs) to obtain a
high level of description. Therefore, we combine the advantage of the low
computational cost of LGCA and its ability to mimic the realistic fluid
dynamics to develop a new animating framework for computer graphics
applications. Up to our knowledge, there are no references using FHP for fluid
animation in Computer Graphics. In this work, we discuss the theoretical
elements of our proposal and present some experimental results.

The paper is organized as follows. Section \ref{Navier} offer a review of CFD
for fluid animation. Section \ref{FHP} describes the FHP model its multiscale
analysis. The experimental results are presented on section \ref{Exper}.
Conclusions are given on Section \ref{Concl}.

\Section{Navier-Stokes for Fluid Animation}\label{Navier}

The majority fluid models in computer graphics follow the Eulerian formulation
of fluid mechanics; that is, the fluid is considered as a continuous system
subjected to Newton's and conservation Laws as well as state equations
connecting the macroscopic variables that define the thermodynamic state of
the fluid: pressure $P,$ density $\rho$ and temperature $T$.

So, the mass conservation, also called continuity equation, is given by
\cite{Hirsch1988}:%

\begin{equation}
\frac{\partial\rho}{\partial t}+\mathbf{\nabla}\cdot(\rho\vec{u})=0
\label{continuidade}%
\end{equation}

The linear momentum conservation equation, also called Navier-Stokes, can be
obtained by applying the third Newton's Law to a volume element $dV$ of
fluid$.$ It can be written as \cite{Hirsch1988}.:
\begin{equation}
\rho\left(  {\frac{\partial\vec{u}}{\partial t}+}\vec{u}{\cdot}\mathbf{\nabla
}\vec{u}\right)  =-\mathbf{\nabla}P+\mathbf{F}+\mu\left(  \mathbf{\nabla}%
^{2}\vec{u}+\frac{1}{3}\mathbf{\nabla}\left(  \mathbf{\nabla\cdot}\vec
{u}\right)  \right)  \label{NavierStokes}%
\end{equation}
where $\mathbf{F}$ is an external force field and $\mu$ is the viscosity of
the fluid. Besides, the equation $\mathbf{\nabla\cdot}\vec{u}=0$ must be added
to model incompressible fluids. Thus, if we combine this equations with
expression (\ref{NavierStokes}) we obtain the Navier-Stokes equations for
incompressible fluids (water, for example):%

\begin{equation}
\rho\left(  {\frac{\partial\vec{u}}{\partial t}+\vec{u}\cdot}\mathbf{\nabla
}\vec{u}\right)  =-\mathbf{\nabla}P+\mathbf{F}+\mu\mathbf{\nabla}^{2}\vec
{u}\mathbf{,} \label{NavierStokes01}%
\end{equation}

\begin{equation}
\mathbf{\nabla\cdot}\vec{u}=0. \label{NavierStokes02}%
\end{equation}

Also, we need an additional equation for the pressure field. This is a state
equation which ties together all of the conservation equations for continuum
fluid dynamics and must be chosen to model the appropriate fluid (\emph{i.e.}
compressible or incompressible). In the case of liquids, the pressure $P$ is
temperature insensitive and can be approximated by $P=P\left(  \rho\right)  $.
Morris in \cite{Morris1997} proposed an expression that have been used for
fluid animation also \cite{muller03}:
\begin{equation}
P=c^{2}\rho\label{press}%
\end{equation}
where $c$ is the speed of sound in this fluid \cite{Schlatter1999}.

Equations (\ref{NavierStokes01})-(\ref{press}) need initial conditions
$\left(  \rho\left(  t=0,x,y,z\right)  ,\vec{u}\left(  t=0,x,y,z\right)
\right)  $. Besides, in practice, fluid domain is a closed subset of the
Euclidean space and thus the behavior of the fluid in the domain boundary -
\textit{boundary conditions} - must be explicitly given. For a fixed rigid
surface $S$, one usual model is the no-sleep boundary condition that can be
written as:%

\begin{equation}
\vec{u}\mathbf{\mid}_{S}=0. \label{eq05}%
\end{equation}

Also, numerical methods should be used to perform the computational simulation
of the fluid because the fluid equations in general do not have analytical
solution. Finite Element (FE) and Finite Difference (FD) are known approaches
in this field. Recently, the Lagrangian \textit{Method of Characteristics
}\cite{stam99,Stam2003} and the meshfree methods of \textit{Smoothed Particle
Hydrodynamics} (\textbf{SPH}) \cite{muller03} and \textit{Moving-Particle
Semi-Implicit} (MPS) \cite{Premoze2003} have been also applied.

If the fluid is temperature sensitive, then an energy conservation law should
be applied. For example, in \cite{Foster-Metaxas-1997} authors develop a
framework for hot turbulent gas animation. The model comprises equations
(\ref{NavierStokes01}),(\ref{NavierStokes02}),(\ref{eq05}) as well as the
following equation for temperature change and the \emph{buoyant} force, respectively:%

\begin{equation}
\frac{\partial T}{\partial t}=\lambda\nabla^{2}T-\nabla\cdot\left(  T\vec
{u}\right)  , \label{energy00}%
\end{equation}

\begin{equation}
\mathbf{F=}-\beta g\left(  T_{0}-T\right)  , \label{energy01}%
\end{equation}
where $\lambda$ is the diffusion coefficient, $T_{0}$ is a reference
temperature and $\beta$ is the coefficient of thermal expansion. The numerical
method used in \cite{Foster-Metaxas-1997} is Finite Difference. This work can
reproduce a hot gas behavior with some realism but has the limitation that the
integration time step is constrained to:%

\begin{equation}
\Delta t<\frac{h}{\left\Vert \vec{u}\right\Vert }, \label{numeric00}%
\end{equation}
where $h$ is the mesh resolution. Besides, the restriction of equation
(\ref{NavierStokes02}) is not suitable for a compressible system like a gas.

Henceforth, after that work, we can find works that: (a) Propose more stable
models to achieve faster simulations; (b) Use truly meshfree Lagrangian
methods; (c) Include realistic behaviors of truly incompressible flow
simulation and interaction of fluids with deformable solids; (d) Use GPU
capabilities in order to achieve faster simulations for interactive
applications; (e) Generate special effects through fluid flows; among others
\cite{SIGGR2004}.

From the viewpoint of fluid models, all the cited works are {top down}
approaches in the sense that the relationships of interest are between
variables that capture the global properties of the system; that is, pressure,
density and temperature. These relationships are expressed in ordinary or
partial differential equations like (\ref{NavierStokes01}).

On the other hand, bottom up models start from a description of local
interactions. These models usually involve algorithmic descriptions of
individuals, \textit{particles} in the case of fluids. Analysis and computer
simulation of bottom up models should produce, as emergent properties, the
global relationships seen in the real world, without these being built into
the model. Thus, there is no need to use a PDEs and numerical methods to
obtain a high level of description.

For Computer Graphics applications, such approach is explored in
\cite{Mark2002CML} for real-time simulation and animation of phenomena
involving convection, reaction-diffusion, and boiling. An extension of
cellular automata known as the coupled map lattice (CML) is used for
simulation. CML represents the state of a dynamic system as continuous values
on a discrete lattice. In \cite{Mark2002CML} the lattice values are stored in
a texture, and pixel-level programming are used to implement simple next-state
computations on lattice nodes and their neighbors. However, Navier-Stokes
models are not considered and CML still uses continuous values for
representations. That is also the case of Lattice Boltzmann models
\cite{HARTING2005}. In this paper we propose the application of an even more
simples model, the FHP one, for fluid simulation. It will be demonstrated how
Navier-Stoke models can be reproduced by this method. FHP is described in the
next section.

\Section{FHP and Navier-Stokes}\label{FHP}

The FHP was introduced by Frisch, Hasslacher and Pomeau \cite{Frisch86} in
1986 and is a model of a two-dimensional fluid. It can be seen as an
abstraction, at a microscopic scale, of a fluid. The FHP model describes the
motion of particles traveling in a discrete space and colliding with each
other. The space is discretized in a hexagonal lattice.

The microdynamics of FHP is given in terms of Boolean variables describing the
occupation numbers at each site of the lattice and at each time step (i.e. the
presence or the absence of a fluid particle). The FHP particles move in
discrete time steps, with a velocity of constant modulus, pointing along one
of the six directions of the lattice. The dynamics is such that no more than
one particle enters the same site at the same time with the same velocity.
This restriction is the \emph{exclusion principle}; it ensures that six
Boolean variables at each lattice site are always enough to represent the microdynamics.

In the absence of collisions, the particles would move in straight lines,
along the direction specified by their velocity vector. The velocity modulus
is such that, in a time step, each particle travels one lattice spacing and
reaches a nearest-neighbor site.

In order to conserve the number of particles and the momentum during each
interaction, only a few configurations lead to a non-trivial collision (i.e. a
collision in which the directions of motion have changed). When exactly two
particles enter the same site with opposite velocities, both of them are
deflected by 60 degrees so that the output of the collision is still a zero
momentum configuration with two particles. When exactly three particles
collide with an angle of 120 degrees between each other, they bounce back to
where they come from (so that the momentum after the collision is zero, as it
was before the collision). Both two- and three-body collisions are necessary
to avoid extra conservation laws. Several variants of the FHP model exist in
the literature \cite{Doolen1990}, including some with rest particles like
models FHP-II and FHP-III. For all other configurations no collision occurs
and the particles go through as if they were transparent to each other.

The full microdynamics of the FHP model can be expressed by evolution
equations for the occupation numbers defined as the number, $n_{i}\left(
\vec{r},t\right)  $, of particle entering site $\vec{r}$ at time $t$ with a
velocity pointing along direction $\vec{c}_{i}$, where $i=1,2,\ldots,6$ labels
the six lattice directions. The numbers $n_{i}$ can be $0$ or $1$.

We also define the time step as $\Delta_{t}$ and the lattice spacing as
$\Delta_{r}$. Thus, the six possible velocities $\vec{v}_{i}$ of the particles
are related to their directions of motion by%
\begin{equation}
\vec{v}_{i}=\frac{\Delta_{r}}{\Delta_{t}}\vec{c}_{i}\text{.}
\label{Equacao 16a - Bastien}%
\end{equation}
Without interactions between particles, the evolution equations for the
$n_{i}$ would be given by%
\begin{equation}
n_{i}\left(  \vec{r}+\Delta_{r}\vec{c}_{i},t+\Delta_{t}\right)  =n_{i}\left(
\vec{r},t\right)  \label{Equacao 17 - Bastien}%
\end{equation}
which express that a particle entering site $\vec{r}$ with velocity along
$\vec{c}_{i}$ will continue in a straight line so that, at next time step, it
will enter site $\vec{r}+\Delta_{r}\vec{c}_{i}$ with the same direction of
motion. However, due to collisions, a particle can be removed from its
original direction or another one can be deflected into direction $\vec{c}%
_{i}$.

For instance, if only $n_{i}$ and $n_{i+3}$ are $1$ at site $\vec{r}$, a
collision occurs and the particle traveling with velocity $\vec{v}_{i}$ will
then move with either velocity $\vec{v}_{i-1}$ or $\vec{v}_{i+1}$, where
$i=1,2,\ldots,6$. The quantity%
\begin{equation}
D_{i}=n_{i}n_{i+3}\left(  1-n_{i+1}\right)  \left(  1-n_{i+2}\right)  \left(
1-n_{i+4}\right)  \left(  1-n_{i+5}\right)  \text{.}
\label{Equacao 18 - Bastien}%
\end{equation}
indicates, when $D_{i}=1$ that such a collision will take place. Therefore
$n_{i}-D_{i}$ is the number of particles left in direction $\vec{c}_{i}$ due
to a two-particle collision along this direction.

\begin{center}%
{\parbox[b]{3.5408in}{\begin{center}
\includegraphics[
height=1.0421in,
width=2.5408in
]%
{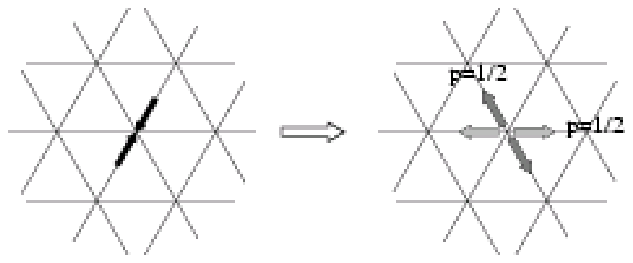}%
\\
Figura 1: The two-body collision in the FHP.
\end{center}}}%

\end{center}

Now, when $n_{i}=0$, a new particle can appear in direction $\vec{c}_{i}$, as
the result of a collision between $n_{i+1}$ and $n_{i+4}$ or a collision
between $n_{i-1}$ e $n_{i+2}$. It is convenient to introduce a random Boolean
variable $q\left(  \vec{r},t\right)  $, which decides whether the particles
are deflected to the right ($q=1$) or to the left ($q=0$), when a two-body
collision takes place. Therefore, the number of particle created in direction
$\vec{c}_{i}$ is%
\begin{equation}
qD_{i-1}+\left(  1-q\right)  D_{i+1}\text{.} \label{Equacao 18a - Bastien}%
\end{equation}
Particles can also be deflected into (or removed from) direction $\vec{c}_{i}$
because of a three-body collision. The quantity which express the occurrence
of a three-body collision with particles $n_{i}$, $n_{i+2}$ and $n_{i+4}$ is%
\begin{equation}
T_{i}=n_{i}n_{i+2}n_{i+4}\left(  1-n_{i+1}\right)  \left(  1-n_{i+3}\right)
\left(  1-n_{i+5}\right)  \label{Equacao 19 - Bastien}%
\end{equation}
As before, the result of a three-body collision is to modify the number of
particles in direction $\vec{c}_{i}$ as
\begin{equation}
n_{i}-T_{i}+T_{i+3}\text{,} \label{Equacao 19a - Bastien}%
\end{equation}
Thus, according to our collision rules, the microdynamics of a LGCA is written
as%
\begin{equation}
n_{i}\left(  \vec{r}+\Delta_{r}\vec{c}_{i},t+\Delta_{t}\right)  =n_{i}\left(
\vec{r},t\right)  +\Omega_{i}\left(  n\left(  \vec{r},t\right)  \right)
\label{Equacao 20 - Bastien}%
\end{equation}
where $\Omega_{i}$ is called the collision term.

For the FHP model, $\Omega_{i}$ is defined so as to reproduce the collisions,
that is%
\begin{equation}
\Omega_{i}=-D_{i}+qD_{i-1}+\left(  1-q\right)  D_{i+1}-T_{i}+T_{i+3}\text{.}
\label{Equacao 21 - Bastien}%
\end{equation}
Using the full expression for $D_{i}$ and $T_{i}$, given by the Equations
(\ref{Equacao 18 - Bastien})-(\ref{Equacao 19 - Bastien}), we obtain,%
\begin{align}
&  \Omega_{i}\label{Equacao 22 - Bastien}\\
&  =-n_{i}n_{i+2}n_{i+4}\left(  1-n_{i+1}\right)  \left(  1-n_{i+3}\right)
\left(  1-n_{i+5}\right) \nonumber\\
&  +n_{i+1}n_{i+3}n_{i+5}\left(  1-n_{i}\right)  \left(  1-n_{i+2}\right)
\left(  1-n_{i+4}\right) \nonumber\\
&  -n_{i}n_{i+3}\left(  1-n_{i+1}\right)  \left(  1-n_{i+2}\right)  \left(
1-n_{i+4}\right)  \left(  1-n_{i+5}\right) \nonumber\\
&  +\left(  1-q\right)  n_{i+1}n_{i+4}\left(  1-n_{i}\right)  \left(
1-n_{i+2}\right)  \left(  1-n_{i+3}\right) \nonumber\\
&  +\left(  1-q\right)  \left(  1-n_{i+5}\right) \nonumber\\
&  +qn_{i+2}n_{i+5}\left(  1-n_{i}\right)  \left(  1-n_{i+1}\right)  \left(
1-n_{i+3}\right)  \left(  1-n_{i+4}\right)  \text{.}\nonumber
\end{align}
These equations are easy to code in a computer and yield a fast and exact
implementation of the model

Until now, we deal with microscopic quantities. However, the physical
quantities of interest are not so much the Boolean variables $n_{i}$ but
macroscopic quantities or average values, such as, for instance, the average
density of particles and the average velocity field at each point of the
system. Theses quantities are defined from the ensemble average $N_{i}\left(
\vec{r},t\right)  =\left\langle n_{i}\left(  \vec{r},t\right)  \right\rangle $
of the microscopic occupation variables. Note that, $N_{i}\left(  \vec
{r},t\right)  $ is also the probability of having a particle entering the site
$\vec{r}$, at time $t$, with velocity%
\[
\vec{v}_{i}=\frac{\Delta_{r}}{\Delta_{t}}\vec{c}_{i}\text{.}%
\]

In general, a LGCA is characterized by the number $z$ of lattice directions
and the spatial dimensionality $d$. In our case $d=2$ and $z=6$. Following the
usual definition of statitical mechanics, the local density of particles is
the sum of the average number of particles traveling along, each direction
$\vec{c}_{i}$%
\begin{equation}
\rho\left(  \vec{r},t\right)  =%
{\displaystyle\sum\limits_{i=0}^{z}}
N_{i}\left(  \vec{r},t\right)  \text{.} \label{Equacao 23 - Bastien}%
\end{equation}
Similarly, the particle current, which is the density $\rho$ times the
velocity field $\vec{u}$, is expressed by.%
\begin{equation}
\rho\left(  \vec{r},t\right)  \vec{u}\left(  \vec{r},t\right)  =%
{\displaystyle\sum\limits_{i=0}^{z}}
\vec{v}_{i}N_{i}\left(  \vec{r},t\right)  \text{.}
\label{Equacao 24 - Bastien}%
\end{equation}
Another quantity which will play an importante role in the up coming
derivation is the momentum tensor $\Pi$ defined as%
\begin{equation}
\Pi_{\alpha\beta}=%
{\displaystyle\sum\limits_{i=0}^{z}}
\vec{v}_{i\alpha}\vec{v}_{i\beta}N_{i}\left(  \vec{r},t\right)
\label{Equacao 25 - Bastien}%
\end{equation}
where the greek indices $\alpha$ and $\beta$ label the $d$ spatial components
of the vectors. The quantity $\Pi$ represents the flux of the $\alpha
-$component of momentum transported along the $\beta-$axis. This term will
contain the pressure contribution and the effects of viscosity.

The starting point to obtain the macroscopic behavior of the CA fluid is to
derive an equation for the $N_{i}^{\prime}s$. Averaging the microdynamics
(\ref{Equacao 20 - Bastien}) yields%
\begin{equation}
N_{i}\left(  \vec{r}+\Delta_{r}\vec{c}_{i},t+\Delta_{t}\right)  -N_{i}\left(
\vec{r},t\right)  =\left\langle \Omega_{i}\left(  n\left(  \vec{r},t\right)
\right)  \right\rangle \label{Equacao 26 - Bastien}%
\end{equation}
where $\Omega_{i}$ is the collision term of the LGCA, under study. It is
important to notice that $\Omega_{i}\left(  n\right)  $ has some generic
properties, namely
\begin{equation}%
{\displaystyle\sum\limits_{i=1}^{z}}
\Omega_{i}=0\text{ \ \ \ \ \ \ \ \ e \ \ \ \ \ \ \ \ }%
{\displaystyle\sum\limits_{i=1}^{z}}
\vec{v}_{i}\Omega_{i}=0 \label{Equacao 27 - Bastien}%
\end{equation}
expressing the fact that particle number and momentum are conserved during the
collision process (the incoming sum of mass or momentum equals the outgoing sum).

The $N_{i}$'s vary between $0$ and $1$ and, at a scale $L>>\Delta_{r}$ e
$T>>\Delta_{t}$, one can expect them to be smooth functions of the space and
time coordinates. Therefore, equation (\ref{Equacao 26 - Bastien}) can be
Taylor expanded up to second order and gives%
\begin{align}
&  \Delta_{r}\left(  \vec{c}_{i}\cdot\nabla\right)  N_{i}\left(  \vec
{r},t\right)  +\Delta_{t}\partial_{t}N_{i}\left(  \vec{r},t\right)
\label{Equacao 28 - Bastien}\\
&  +\frac{1}{2}\left(  \Delta_{r}\right)  ^{2}\left(  \vec{c}_{i}\cdot
\nabla\right)  ^{2}N_{i}\left(  \vec{r},t\right)  +\Delta_{r}\Delta_{t}\left(
\vec{c}_{i}\cdot\nabla\right)  \partial_{t}N_{i}\left(  \vec{r},t\right)
\nonumber\\
&  +\frac{1}{2}\left(  \Delta_{t}\right)  ^{2}\left(  \partial_{t}\right)
^{2}N_{i}\left(  \vec{r},t\right)  =\left\langle \Omega_{i}\left(  n\left(
\vec{r},t\right)  \right)  \right\rangle \text{.}\nonumber
\end{align}
where $\left(  \partial_{t}\right)  ^{2}$ is the second derivative in respect
to the time parameter $t$.

At a macroscopic scale $L>>\Delta_{r}$, following the procedure of the
so-called multiscale expansion \cite{Piasecki1997}, we introduce a new space
variable $\vec{r}_{1}$ such that%
\begin{equation}
\vec{r}_{1}=\epsilon\partial_{\vec{r}_{1}}\text{ \ \ \ \ e \ \ \ \ \ }%
\partial_{r}=\epsilon\partial_{\vec{r}_{1}} \label{Equacao 29 - Bastien}%
\end{equation}
with $\epsilon<<1$. We also introduce the extra time variables $t_{1}=\epsilon
t$ and $t_{2}=\epsilon^{2} t$, as well as new functions $N_{i}^{\epsilon}$
depending on $\vec{r}_{1}$, $t_{1}$ and $t_{2}$, $N_{i}^{\epsilon}%
=N_{i}^{\epsilon}\left(  t_{1},t_{2},\vec{r}_{1}\right)  $ and substitute into
equation (\ref{Equacao 28 - Bastien})%
\begin{equation}
N_{i}\rightarrow N_{i}^{\epsilon}\text{ \ \ \ \ \ }\partial_{t}\rightarrow
\epsilon\partial_{t_{1}}+\epsilon^{2}\partial_{t_{2}}\text{\ \ \ \ \ \ \ \ }%
\partial_{r}\rightarrow\epsilon\partial_{\vec{r}_{1}}
\label{Equacao 30 - Bastien}%
\end{equation}
together with the corresponding expressions for the second order derivatives.
Then obtain new equations for the new functions $N_{i}^{\epsilon}$. Thus, we
may write \cite{Piasecki1997},
\begin{equation}
N_{i}^{\epsilon}=N_{i}^{\left(  0\right)  }+\epsilon N_{i}^{\left(  1\right)
}+\epsilon^{2}N_{i}^{\left(  2\right)  }+\cdots\label{Equacao 31 - Bastien}%
\end{equation}

The Chapman-Enskog method is the standard procedure used in statistical
mechanics to solve an equation like (\ref{Equacao 28 - Bastien}) with a
perturbation parameter $\epsilon$. Assuming that $\left\langle \Omega
_{i}\left(  n\right)  \right\rangle $ can be factorized into $\Omega
_{i}\left(  N\right)  $, we write the contributions of each order in
$\epsilon$. According to multiscale expansion (\ref{Equacao 31 - Bastien)},
the right-hand side of (\ref{Equacao 28 - Bastien}) reads%

\begin{equation}
\Omega_{i}\left(  N\right)  =\Omega_{i}\left(  N^{\left(  0\right)  }\right)
+\epsilon%
{\displaystyle\sum\limits_{j=1}^{z}}
\left(  \frac{\partial\Omega_{i}\left(  N^{\left(  0\right)  }\right)
}{\partial N_{j}}\right)  N_{j}^{\left(  1\right)  }+\mathcal{O}\left(
\epsilon^{2}\right)  \label{Equacao 32a - Bastien}%
\end{equation}
Using expressions (\ref{Equacao 29 - Bastien})-(\ref{Equacao 31 - Bastien}) in
the left-hand side of (\ref{Equacao 28 - Bastien}) and comparing the terms of
the same order in $\epsilon$\ in the equation (\ref{Equacao 32a - Bastien}), yields%

\begin{equation}
O\left(  \epsilon^{0}\right)  :\Omega_{i}\left(  N^{\left(  0\right)
}\right)  =0 \label{Equacao 33 - Bastien}%
\end{equation}
and
\begin{align}
O\left(  \epsilon^{1}\right)   &  :\partial_{1\alpha}v_{i\alpha}N_{i}^{\left(
0\right)  }+\partial_{t_{1}}N_{i}^{\left(  0\right)  }%
\label{Equacao 34 - Bastien}\\
&  =\frac{1}{\Delta_{t}}%
{\displaystyle\sum\limits_{j=1}^{z}}
\left(  \frac{\partial\Omega_{i}\left(  N^{\left(  0\right)  }\right)
}{\partial N_{j}}\right)  N_{j}^{\left(  1\right)  }\nonumber
\end{align}
where the subscript $1$ in spatial derivatives (e.g. $\partial_{1\alpha}$)
indicates a differential operator expressed in the variable $\vec{r}_{1}$ and
$\frac{\Delta_{r}}{\Delta_{t}}\left(  \vec{c}_{i}\cdot\nabla_{r_{1}}\right)
=\partial_{1\alpha}v_{i\alpha}$, from equation (\ref{Equacao 16a - Bastien}).

We also impose the extra conditions that the macroscopic quantities $\rho$ and
$\rho\vec{u}$ are entirely given by the zero order of expansion
(\ref{Equacao 31 - Bastien})%
\begin{equation}
\rho=%
{\displaystyle\sum\limits_{i=1}^{z}}
N_{i}^{\left(  0\right)  }\text{ \ \ and \ \ \ \ }\rho\vec{u}=%
{\displaystyle\sum\limits_{i=1}^{z}}
\vec{v}_{i}N_{i}^{\left(  0\right)  } \label{Equacao 39 - Bastien}%
\end{equation}
and therefore%
\begin{equation}%
{\displaystyle\sum\limits_{i=1}^{z}}
N_{i}^{\left(  l\right)  }=0\text{ \ \ \ \ \ and \ \ \ \ }%
{\displaystyle\sum\limits_{i=1}^{z}}
\vec{v}_{i}N_{i}^{\left(  l\right)  }=0\text{, \ \ \ \ for }l\geq1
\label{Equacao 40 - Bastien}%
\end{equation}

Thus, following the Chapman-Enskog method we can obtain \cite{Frisch87}, from
equation (\ref{Equacao 28 - Bastien}), the following result at order
$\epsilon$%
\begin{equation}
\partial_{t_{1}}\rho+\operatorname*{div}\nolimits_{1}\rho u=0
\label{Equacao 41 - Bastien}%
\end{equation}
and
\begin{equation}
\partial_{t_{1}}\rho u_{\alpha}+\partial_{1\beta}\Pi_{\alpha\beta}^{\left(
0\right)  }=0 \label{Equacao 42 - Bastien}%
\end{equation}
On the other hand, if we consider the terms of order $\epsilon^{2}$ and using
the relations (\ref{Equacao 41 - Bastien}) and (\ref{Equacao 42 - Bastien}) to
simplify, we have%
\begin{equation}
\partial_{t_{2}}\rho u_{a}+\partial_{1\beta}\left[  \Pi_{\alpha\beta}^{\left(
1\right)  }+\frac{\Delta_{t}}{2}\left(  \partial_{t_{1}}\Pi_{\alpha\beta
}^{\left(  0\right)  }+\partial_{1\gamma}S_{\alpha\beta\gamma}^{\left(
0\right)  }\right)  \right]  =0 \label{Equacao 49 - Bastien}%
\end{equation}
The last equation contains the dissipative contributions to the Euler equation
(\ref{Equacao 42 - Bastien}). The first contribution is $\Pi_{\alpha\beta
}^{\left(  1\right)  }$ which is the dissipative part of the momentum tensor.
The second part, namely $\frac{\Delta_{t}}{2}\left(  \partial_{t_{1}}%
\Pi_{\alpha\beta}^{\left(  0\right)  }+\partial_{1\gamma}S_{\alpha\beta\gamma
}^{\left(  0\right)  }\right)  $ comes from the second order terms of the
Taylor expansion of the discrete Boltzmann equation. These terms account for
the discreteness of the lattice and have no counterpart in standard
hydrodynamics. As we shall see, they will lead to the so-called lattice
viscosity. The order $\epsilon$ e $\epsilon^{2}$ can be grouped together to
give the general equations governing our system. Summing equations
(\ref{Equacao 41 - Bastien}) and (\ref{Equacao 49 - Bastien}) with the
appropriate power of $\epsilon$ as factor and we obtain the continuity
equation (see expression (\ref{continuidade}):%
\begin{equation}
\partial_{t}\rho+\operatorname*{div}\rho\vec{u}=0 \label{Equacao 51 - Bastien}%
\end{equation}
Similarly, equation (\ref{Equacao 42 - Bastien}) and
(\ref{Equacao 49 - Bastien}) yields \cite{Frisch87}%
\begin{equation}
\partial_{t}\rho u_{a}+\frac{\partial}{\partial_{r_{\beta}}}\left[
\Pi_{\alpha\beta}+\frac{\Delta_{t}}{2}\left(  \epsilon\partial_{t_{1}}%
\Pi_{\alpha\beta}^{\left(  0\right)  }+\frac{\partial}{\partial_{r_{\gamma}}%
}S_{\alpha\beta\gamma}^{\left(  0\right)  }\right)  \right]  =0
\label{Equacao 50 - Bastien}%
\end{equation}

We now turn to the problem of solving equation (\ref{Equacao 33 - Bastien})
together with conditions (\ref{Equacao 39 - Bastien}) in order to find
$N_{i}^{\left(  0\right)  }$ as functions of $\rho$ and $\rho\vec{u}$. The
solutions $N_{i}^{\left(  0\right)  }$ which make the collision term $\Omega$
vanish are known as the local equilibrium solutions. Physically, they
correspond to a situation where the rate of each type of collision
equilibrates. Since the collision time $\Delta_{t}$ is much smaller than the
macroscopic observation time, it is reasonable to expect, in first
approximation that an equilibrium is reached locally.

Provided that the collision behaves reasonably, it is found \cite{Frisch87}
that the generic solution is%
\begin{equation}
N_{i}^{\left(  0\right)  }=\frac{1}{1+\exp\left(  -A-\vec{B}\cdot\vec{v}%
_{i}\right)  } \label{Equacao 52 - Bastien}%
\end{equation}
This expression has the form of a Fermi-Dirac distribution. This is a
consequence of the exclusion principle we have imposed in the cellular
automata rule\ (no more than one particle per site and direction). This form
is explicitly obtained for the FHP model by assuming that the rate of direct
and inverse collisions are equal. The quantities $A$ e $\vec{B}$ are functions
of the density $\rho$ and the velocity field $\vec{u}$ and are to be
determined according to equations (\ref{Equacao 39 - Bastien}). In order to
carry out this calculation, $N_{i}^{\left(  0\right)  }$ is Taylor expanded up
to second order in the velocity field $\vec{u}$. One obtains
\cite{Chopard1998}%
\begin{equation}
N_{i}^{\left(  0\right)  }=a\rho+\frac{b\rho}{v^{2}}\vec{v}_{i}\cdot\vec
{u}+\frac{\rho G\left(  \rho\right)  }{v^{4}}Q_{i\alpha\beta}u_{\alpha
}u_{\beta} \label{Equacao 53 - Bastien}%
\end{equation}
where $\alpha,\beta,\gamma$ are summed over the spacial coordinates, e.g.
$\alpha,\beta,\gamma\in\left\{  1,\ldots,d\right\}  $, $v=\frac{\Delta_{r}%
}{\Delta_{t}}$, $a=\frac{1}{z}$, $b=\frac{d}{z}$ and%
\begin{equation}
Q_{i\alpha\beta}=v_{i\alpha}v_{i\beta}-\frac{v^{2}}{d}\delta_{\alpha\beta}
\label{Equacao 54 - Bastien}%
\end{equation}

The function $G$ is obtained from the fact that $N_{i}^{\left(  0\right)  }$
is the Taylor expansion of a Fermi-Dirac distribution. For FHP, it is found
\cite{Chopard1998,Frisch87}%
\[
G\left(  \rho\right)  =\frac{2}{3}\frac{\left(  3-\rho\right)  }{\left(
6-\rho\right)  }%
\]

We may now compute the local equilibrium part of the momentum tensor,
$\Pi_{\alpha\beta}^{\left(  0\right)  }$ and then obtain the pressure term
\begin{equation}
p=aC_{2}v^{2}\rho-\left[  \frac{C_{2}}{d}-C_{4}\right]  \rho G\left(
\rho\right)  u^{2} \label{Equacao 60 - Bastien}%
\end{equation}
where $C_{2}=\frac{z}{d}$.

We can see \cite{Frisch87} that the lattice viscosity is given by%
\begin{align*}
\nu_{lattice}  &  =-C_{4}b\frac{\Delta_{t}v^{2}}{2}=-\frac{z}{d\left(
d+2\right)  }\frac{d}{z}\frac{\Delta_{t}}{2}v^{2}\\
&  =\frac{-\Delta_{t}}{2\left(  d+2\right)  }v^{2}%
\end{align*}
The usual contribution to viscosity is due to the collision between the fluid
particles is given by \cite{Frisch87}%
\[
\nu_{coll}=\Delta_{t}v^{2}\frac{bC_{4}}{\Lambda}%
\]
where $-\Lambda$ is given by $-\Lambda=2s\left(  1-s\right)  ^{3}$ where
$s=\frac{\rho}{6}$

Therefore, the Navier-Stokes equation reads%
\begin{equation}
\partial_{t}\vec{u}+2C_{4}G\left(  \rho\right)  \left(  \vec{u}\cdot
\nabla\right)  \vec{u}=-\frac{1}{\rho}\nabla p+\nu\nabla^{2}\vec{u}
\label{Equacao 74 - Bastien}%
\end{equation}
where%
\begin{equation}
\nu=\Delta_{t}v^{2}bC_{4}\left(  \frac{1}{\Lambda}-\frac{1}{2}\right)
=\frac{\Delta_{t}v^{2}}{d+2}\left(  \frac{1}{\Lambda}-\frac{1}{2}\right)
\label{Equacao 75 - Bastien}%
\end{equation}
is the kinematic viscosity of our discrete fluid.

\Section{Experimental Results}\label{Exper}

In this section we describe some experiments with FHP for bidimensional fluid
simulation. Firstly, we highlight the simplicity of creating new
configurations. Figure 2 shows an initial configuration with zero density in
the middle of the system. It is not required any extra mathematical machinery
to deal with such density discontinuity because system rules do not undergo
modifications. Figures 1 were generated with 80.000 particles with position
and velocity directions randomly distributed. The lattice resolution is 100 by
100 points.

The density distribution at time 10 and 25 (Figures 3 and 4) show an
interesting pattern near the front of the discontinuity. The evolution for
time 50 is even more interesting (Figure 5). If we want to predict such
effects, we need to consider Navier-Stokes equations. However, if the aim is
to explore the visual effect, we can just simulate and take the desired result
at its time. As expected, the system evolves towards a configuration in
thermodynamic equilibrium (or maximum entropy \cite{Wolfram1996}). Figure 6
shows such state. From the macroscopic viewpoint, the fluid achieves a static
configuration in which the macroscopic velocity $\vec{u}$ is null everywhere.
If we decrease particle density, the pattern obtained is basically the same,
as we can verify through Figure 7.

The configuration of \ pictured on Figure 2 can be generalized by an initial
density with a disconnected zero set. Figure 10-a pictures such example. We
get an interesting pattern formation presented on Figure 10-b. These patterns
evolve to the "S" formations pictured on Figure 11.

Besides, we can take advantage of the simplicity of the model for changing
boundary. For a LGCA, there is no need to re-build the lattice. It is just a
matter of finding the boundary cells of the lattice and apply the proper
collision rules for particles entering the corresponding sites. Next, we show
the tests using a homogeneous particles distribution with velocity in the
horizontal direction. It is interesting to observe the patterns at the right
hand side of the Figure 9. Particles that collide with the domain boundary
also will collide with the insident particles which increases the density nearby.

\Section{Conclusions}\label{Concl}

In this paper we propose the FHP model for fluid modeling in computer graphics
applications. We discuss the theoretical elements of our proposal and discuss
some experimental results. Further works are the incorporation of external
forces and model two-fase systems for visual effects generation.

\bibliographystyle{latex8}
\bibliography{Sibgrapi2005a}

\newpage
\clearpage\onecolumn\vfill%
\raisebox{-1.5108in}{\parbox[b]{1.4659in}{\begin{center}
\includegraphics[
height=1.4736in,
width=1.4659in
]%
{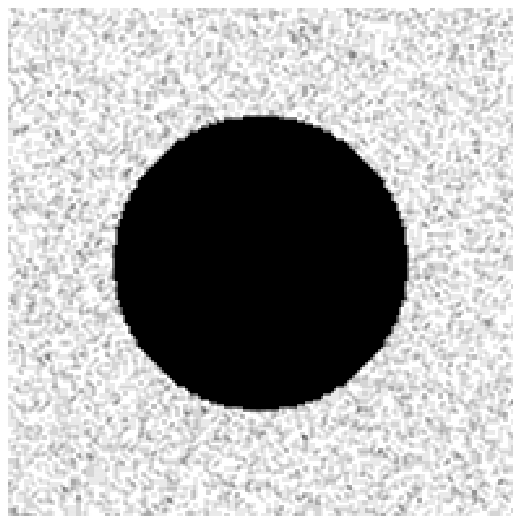}%
\\
Figure 2: Initial configuration with 80000
\end{center}}}
\ \ \ \
\raisebox{-1.5108in}{\parbox[b]{1.497in}{\begin{center}
\includegraphics[
height=1.4901in,
width=1.497in
]%
{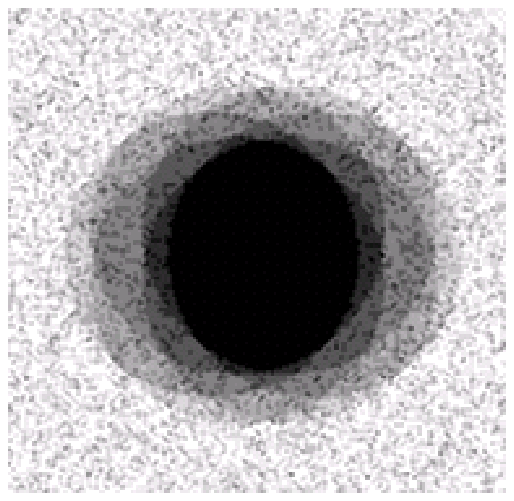}%
\\
Figure 3: Evolution after 10 steps
\end{center}}}
\ \ \ \
\raisebox{-1.5108in}{\parbox[b]{1.497in}{\begin{center}
\includegraphics[
height=1.5126in,
width=1.497in
]%
{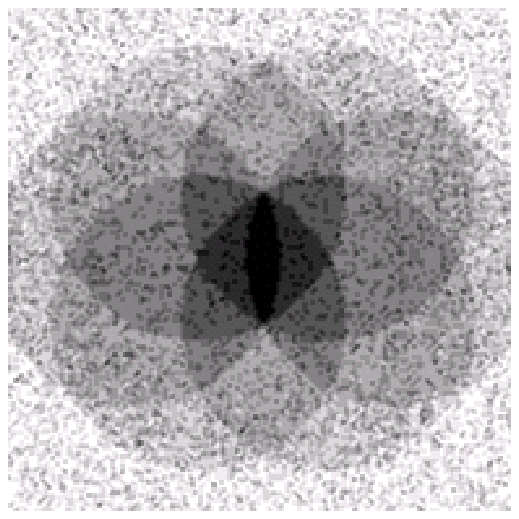}%
\\
Figure 4: Evolution after 25 steps
\end{center}}}
\ \ \ \
\raisebox{-1.4996in}{\parbox[b]{1.497in}{\begin{center}
\includegraphics[
height=1.5281in,
width=1.497in
]%
{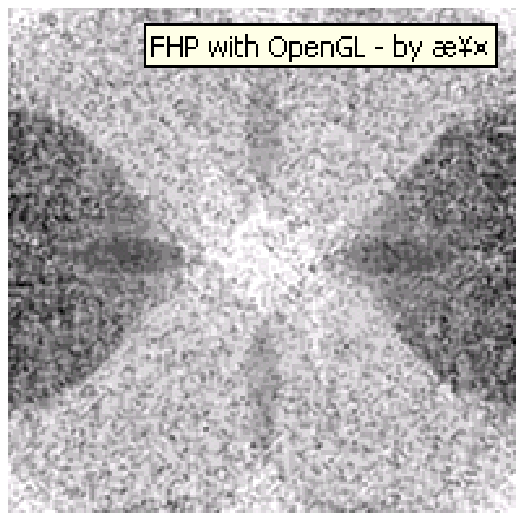}%
\\
Figure 5: Evolution after 50 steps
\end{center}}}
%

\raisebox{-1.5316in}{\parbox[b]{1.4866in}{\begin{center}
\includegraphics[
height=1.4935in,
width=1.4866in
]%
{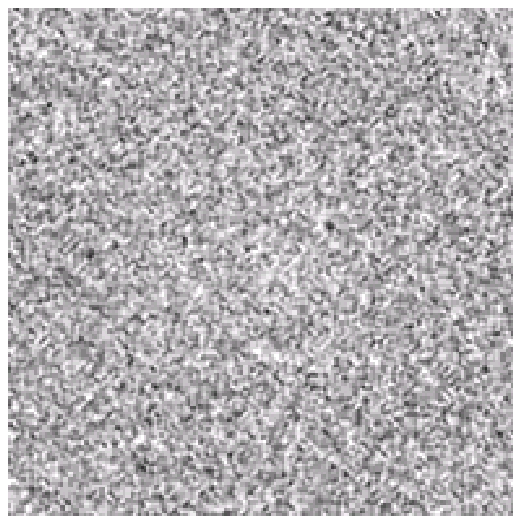}%
\\
Figure 6: Evolution after 300 eteps
\end{center}}}
\ \ \ \
\raisebox{-1.542in}{\parbox[b]{1.5082in}{\begin{center}
\includegraphics[
height=1.4996in,
width=1.5082in
]%
{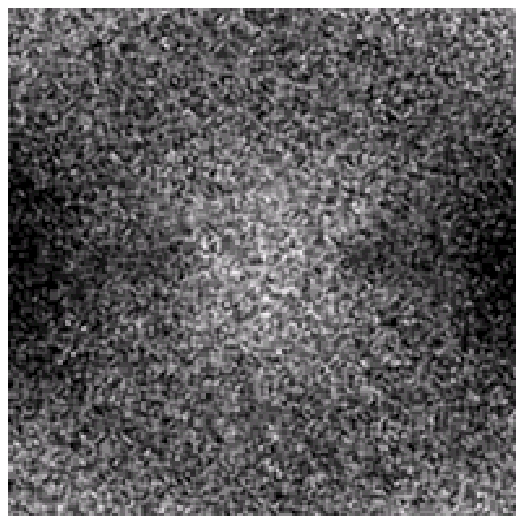}%
\\
Figure 7: 40000 particles after 50 steps
\end{center}}}
\ \ \ \
\raisebox{-1.5627in}{\parbox[b]{1.5186in}{\begin{center}
\includegraphics[
height=1.5186in,
width=1.5186in
]%
{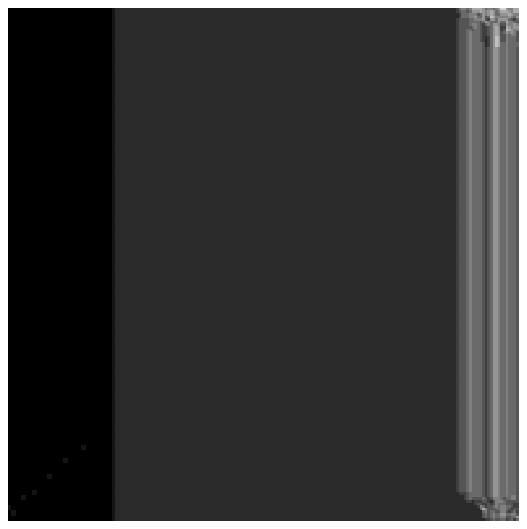}%
\\
Figure 8: Horizontal velocity pattern
\end{center}}}
\ \ \ \
\raisebox{-1.5731in}{\parbox[b]{1.5186in}{\begin{center}
\includegraphics[
height=1.5186in,
width=1.5186in
]%
{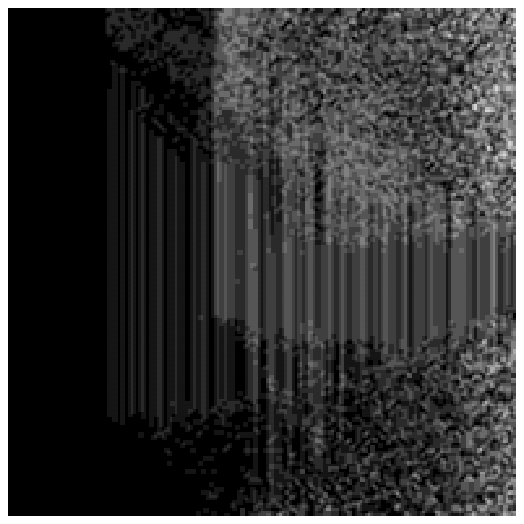}%
\\
Figure 9: Evolution after 100 steps of Figure 7
\end{center}}}
%

\raisebox{--0.1669in}{\parbox[b]{3.1739in}{\begin{center}
\includegraphics[
height=1.6431in,
width=3.1739in
]%
{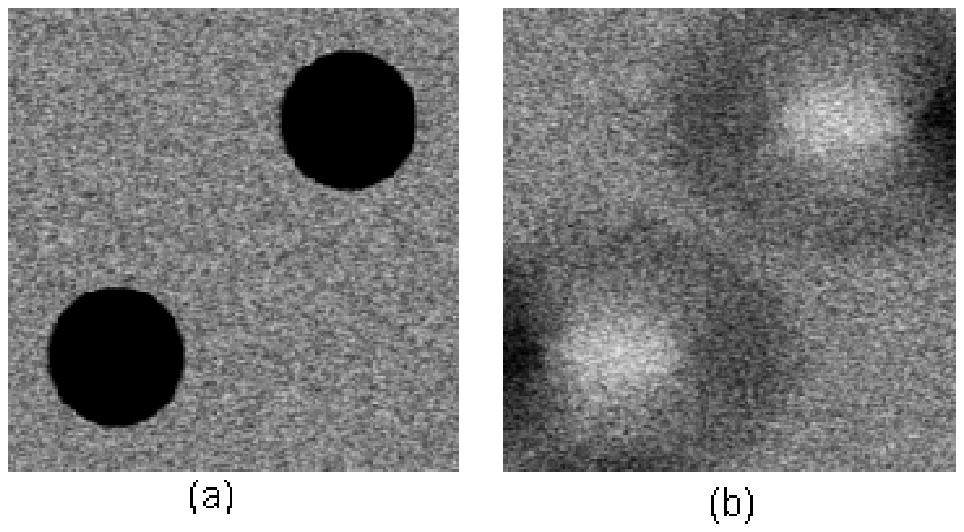}%
\\
Figure 10: (a) Initial configuration. (b) Trasient pattern formation.
\end{center}}}
%

\raisebox{-0.0519in}{\parbox[b]{6.7421in}{\begin{center}
\includegraphics[
height=1.4961in,
width=6.7421in
]%
{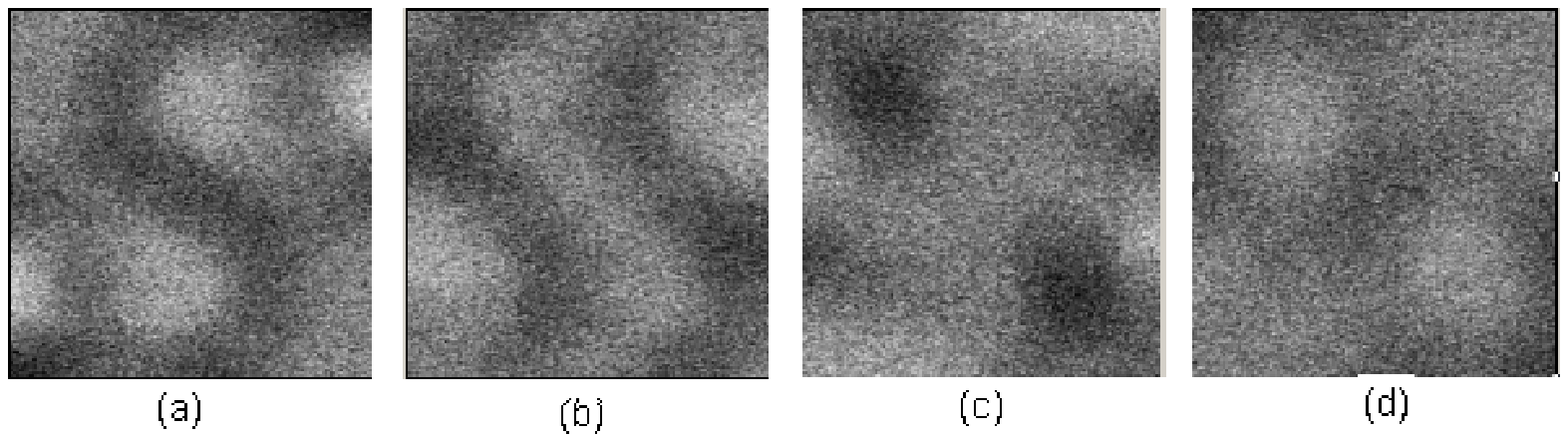}%
\\
Figure 11: Evolution of the configuration pictured on Figure 9-a.
\end{center}}}

\end{document}